\documentclass{JHEP3}
\usepackage{graphicx}

\catcode`@=11
\newcount\countdp \newcount\countwd \newcount\countht 
\@ifundefined{pdfoutput}{
\def\rgboo#1{}
\def\postscript#1{\special{" #1}}		
\postscript{
        /bd {bind def} bind def
        /fsd {findfont exch scalefont def} bd
        /sms {setfont moveto show} bd
        /ms {moveto show} bd
        /pdfmark where          
        {pop} {userdict /pdfmark /cleartomark load put} ifelse
        [ /PageMode /UseOutlines                
        /DOCVIEW pdfmark}
\def\hhref#1#2{%
        \hskip-.25em\setbox0=\hbox{#1}%
                \countdp=\dp0 \countwd=\wd0 \countht=\ht0%
                \divide\countdp by65536 \divide\countwd by65536%
                        \divide\countht by65536%
                \advance\countdp by1 \advance\countwd by1%
                        \advance\countht by1%
                \def\linkdp{\the\countdp} \def\linkwd{\the\countwd}%
                        \def\linkht{\the\countht}%
        \postscript{
                [ /Rect [ -1.5 -\linkdp.0 0\linkwd.0 0\linkht.5 ] 
                /Border [ 0 0 0 ]
                /Action << /Subtype /URI /URI (#1) >>
                /Subtype /Link
                /ANN pdfmark}{\rgb{.1 .1 1}{#2}}}
}{
\def\rgboo#1{\pdfliteral{#1 rg #1 RG}}
\def\hhref#1#2{%
        \noindent\pdfstartlink user
                {/Subtype /Link
                /Border [ 0 0 0 ]
                /A << /S /URI /URI (#1) >>}{\rgb{.1 .1 1}{#2}}%
        \pdfendlink}
}
\def\rgbo#1#2{\rgboo{#1}#2\rgboo{0 0 0}}
\def\rgb#1#2{\mark{#1}\rgbo{#1}{#2}\mark{0 0 0}}

\def\@spires#1{\hhref{http://www-spires.slac.stanford.edu/spires/find/hep/www?j=#1}} 
\begingroup
\catcode`\ =\active\catcode`,=\active\global
\def\@@keywords#1{\gdef\@keywords{\noindent{\scshape\keywordsname}
		\bgroup\def, {+}\def {_}
		\hhref{http://jhep.sissa.it/stdsearch}%
						{\let,\@comma\let \ #1}.
		\egroup}\egroup\global\@keywordstrue}%
\endgroup
\catcode`@=12
\catcode`\%=12
\renewcommand\jhep[3]  {\hhref{http://jhep.sissa.it/stdsearch?paper=#1
		{{\it J. High Energy Phys.\ }{\bf #1} (#2) #3}}
\catcode`\%=14
\renewcommand{\hepth}[1]{\hhref{http://xxx.lanl.gov/abs/hep-th/#1}{\tt hep-th/#1}}
\renewcommand{\hepph}[1]{\hhref{http://xxx.lanl.gov/abs/hep-ph/#1}{\tt hep-ph/#1}}
\renewcommand{\grqc}[1]{\hhref{http://xxx.lanl.gov/abs/gr-qc/#1}{\tt gr-qc/#1}}
\renewcommand{\astroph}[1]{\hhref{http://xxx.lanl.gov/abs/astro-ph/#1}{\tt astro-ph/#1}}
\renewcommand\email[1]{{\tt\hhref{mailto:#1}{#1}}}

\renewcommand{\theequation}{\thesection.\arabic{equation}}
\newcommand{\be}{\begin{equation}}
\newcommand{\ee}{\end{equation}}
\newcommand{\beqy}{\begin{eqnarray}}
\newcommand{\eeqy}{\end{eqnarray}}
\newcommand{\p}{\partial}
\newcommand{\hp}{\widehat{\p}}
\newcommand{\ov}{\overline}
\newcommand{\da}{^{\dagger}}
\newcommand{\w}{\wedge}
\newcommand{\st}{\stackrel}
\newcommand{\mb}{\mbox}
\newcommand{\mx}{\mbox}
\newcommand{\mt}{\mathtt}
\newcommand{\dt}{\mathtt{d}}
\newcommand{\al}{\alpha}
\newcommand{\bb}{\beta}
\newcommand{\ga}{\gamma}
\newcommand{\Ga}{\Gamma}
\newcommand{\te}{\theta}
\newcommand{\Te}{\Theta}
\newcommand{\de}{\delta}
\newcommand{\De}{\Delta}
\newcommand{\ka}{\kappa}
\newcommand{\et}{\tilde{e}}
\newcommand{\ze}{\zeta}
\newcommand{\s}{\sigma}
\newcommand{\e}{\epsilon}
\newcommand{\om}{\omega}
\newcommand{\Om}{\Omega}
\newcommand{\la}{\lambda}
\newcommand{\La}{\Lambda}
\newcommand{\n}{\nabla}
\newcommand{\hn}{\widehat{\nabla}}
\newcommand{\hph}{\widehat{\phi}}
\newcommand{\ah}{\widehat{a}}
\newcommand{\bh}{\widehat{b}}
\newcommand{\ch}{\widehat{c}}
\newcommand{\ddh}{\widehat{d}}
\newcommand{\eh}{\widehat{e}}
\newcommand{\gh}{\widehat{g}}
\newcommand{\ph}{\widehat{p}}
\newcommand{\qh}{\widehat{q}}
\newcommand{\mh}{\widehat{m}}
\newcommand{\nh}{\widehat{n}}
\newcommand{\Dh}{\widehat{D}}
\newcommand{\stu}{\st{\textvisiblespace}}
\newcommand{\au}{\stu{a}}
\newcommand{\bu}{\stu{b}}
\newcommand{\cu}{\stu{c}}
\newcommand{\du}{\stu{d}}
\newcommand{\eu}{\stu{e}}
\newcommand{\mmu}{\stu{m}}
\newcommand{\nnu}{\stu{n}}
\newcommand{\pu}{\stu{p}}
\newcommand{\Du}{\stu{D}}
\newcommand{\sto}{\st{\circ}}
\newcommand{\as}{\st{\circ}{a}}
\newcommand{\bs}{\st{\circ}{b}}
\newcommand{\cs}{\st{\circ}{c}}
\newcommand{\ds}{\st{\circ}{d}}
\newcommand{\es}{\st{\circ}{e}}
\newcommand{\ms}{\st{\circ}{m}}
\newcommand{\ns}{\st{\circ}{n}}
\newcommand{\ps}{\st{\circ}{p}}
\newcommand{\Ds}{\st{\circ}{D}}
\newcommand{\sts}{\st{s}}
\newcommand{\sth}{\st{\heartsuit}}
\newcommand{\stp}{\st{\perp}}
\newcommand{\std}{\st{\diamondsuit}}
\newcommand{\ad}{\dot{a}}
\newcommand{\bd}{\st{s}{b}}
\newcommand{\cd}{\st{s}{c}}
\newcommand{\gd}{\st{s}{g}}
\newcommand{\dd}{\st{s}{d}}
\newcommand{\Dd}{\st{s}{D}}
\newcommand{\ed}{\st{s}{e}}
\newcommand{\fd}{\st{s}{f}}
\newcommand{\zd}{\st{s}{\xi}}
\newcommand{\md}{\st{s}{m}}
\newcommand{\nd}{\st{s}{n}}
\newcommand{\stc}{\st{c}}
\newcommand{\az}{\st{c}{a}}
\newcommand{\bz}{\st{c}{b}}
\newcommand{\cz}{\st{c}{c}}
\newcommand{\dz}{\st{c}{d}}
\newcommand{\Dz}{\st{c}{D}}
\newcommand{\ez}{\st{c}{e}}
\newcommand{\fz}{\st{c}{f}}
\newcommand{\nz}{\st{c}{n}}
\newcommand{\mz}{\st{c}{m}}
\newcommand{\tb}{\overline{\theta}}
\newcommand{\ti}{\widetilde}

\newcommand{\cE}{{\cal E}}
\newcommand{\cN}{{\cal N}}
\newcommand{\cO}{{\cal O}}

\newcommand{\hb}{\bar{h}}
\newcommand{\sqw}{\sqrt{w\over 2}\ }

\newcommand{\2}{\frac{1}{2}}
\newcommand{\3}{\frac{1}{3}}
\newcommand{\4}{\frac{1}{4}}
\newcommand{\8}{\frac{1}{8}}
\newcommand{\6}{\frac{1}{16}}
\newcommand{\stwo}{\sqrt{2}}
\newcommand{\LT}{\left[}
\newcommand{\RT}{\right]}
\newcommand{\LF}{\left(}
\newcommand{\RF}{\right)}
\newcommand{\ra}{\rightarrow}
\newcommand{\Ra}{\Rightarrow}
\newcommand{\im}{\Longleftrightarrow}
\newcommand{\hs}{\hspace{5mm}}
\newcommand{\x}{\star}
\newcommand{\Delt}{\p^{\star}}
\newcommand{\vs}{\vspace{5mm}\\}
\newcommand{\ie}{{\it i.e. }}
\preprint{YITP-SB-06-46}
\title{Non-perturbative Gravity, Hagedorn Bounce \& CMB }

\author{Tirthabir Biswas \& Robert Brandenberger\\
McGill University, 3600 University Ave., Montr\'eal, Qu\'ebec, Canada\\
\email{biswas@hep.physics.mcgill.ca}, \email{rhb@hep.physics.mcgill.ca}}
\author{Anupam Mazumdar\\
NORDITA, Blegdamsvej-17, DK-2100, Copenhagen, Denmark\\
\email{anupamm@nordita.dk}}
\author{Warren Siegel\\
C.N. Yang Institute for Theoretical Physics, State University of New 
York,\\
Stony Brook, NY 11794-3840, USA\\
\email{siegel@insti.physics.sunysb.edu}}

\abstract{In~\cite{BMS} it was shown how  non-perturbative corrections 
to gravity can resolve the big bang singularity, leading to a bouncing universe.
Depending on the scale of the non-perturbative corrections, the temperature
at the bounce may be close to or higher than the Hagedorn temperature. If
matter is made up of strings, then massive string states will be excited
near the bounce, and the bounce will occur inside (or at the onset of) the 
Hagedorn phase for string matter. As we discuss in this paper, in this case 
cosmological fluctuations can 
be generated via the string gas mechanism recently proposed in \cite{NBV}. 
In fact, the model discussed here demonstrates explicitly that it is 
possible to realize the assumptions made in \cite{NBV} in the context of a 
concrete set of dynamical background equations.
We also calculate the spectral tilt of thermodynamic stringy fluctuations 
generated in the Hagedorn regime in this bouncing universe scenario. 
Generally we find a scale-invariant spectrum with a red tilt which is very 
small but does not vanish.} 

\begin{document}

\section{Introduction}
Recently, a new structure formation scenario has been put forward in 
\cite{NBV,BNPV} (see also \cite{Ali,BNPV2} for more in depth discussions). 
In these works it was
{\it shown that string thermodynamic fluctuations in a quasi-static
primordial Hagedorn phase \cite{BV} (during which the temperature hovers
near its limiting value in string theory, namely the Hagedorn 
temperature, $T_H$, \cite{Hagedorn}, the metric is almost static, and hence the Hubble radius
is almost infinite) can lead to a scale-invariant
spectrum of metric fluctuations.} To obtain this result, several criteria
for the background cosmology need to be satisfied (see \cite{Betal}).
First of all, the background equations must indeed admit a quasi-static
(loitering) solution.
Next, our three large spatial dimensions must be compact. It is
under this condition that \cite{Deo2} the heat capacity $C_V$ as a 
function of radius $r$ scales as $r^2$. Thirdly, thermal equilibrium must be 
present over a scale larger than $1$mm \footnote{This scale is obtained
by evolving our present Hubble radius back to temperatures of the order
of the scale of Grand Unification using the equations of Standard Cosmology.}
during the stage of the early universe
when the fluctuations are generated. Since the scale of thermal
equilibrium is bounded from above by the Hubble radius, it follows that
in order to have thermal equilibrium on the required scale, 
the background cosmology should have a quasi-static phase. Finally,
the dilaton velocity needs to be negligible during the time interval
when fluctuations are generated. 

It is not easy to satisfy all of the conditions required for the
mechanism proposed in \cite{NBV} to work. In the context of a dilaton gravity
background, the dynamics of the dilaton is important. If the 
dilaton has not obtained a large mass and a fixed
vacuum expectation value (VEV) at a high scale, then it will be rolling
towards weak coupling at early times. This will lead to \cite{TV,Ven} a
phase in which the string frame metric is static, and thus the
string frame Hubble radius will tend to infinity, i.e. $H\approx 0$. 
However, the large dilaton velocity during this phase changes the
spectrum of fluctuations coming from the stringy Hagedorn phase 
\cite{Betal,KKLM}. In addition, the duration of the Hagedorn phase is too small for the establishment of thermal equilibrium on
sufficiently large scales \cite{Betal} (also see \cite{frey}). {\it These problems can be
overcome in a model which has fixed dilaton and admits a bouncing solution
\footnote{A bouncing cosmology automatically implies a phase of
acceleration. However, in general this phase will be short (of
the order or smaller than one Hubble expansion time), and will
thus not lead to a long period of cosmological inflation.}. 
Such a bouncing solution for the Einstein frame metric turns out
to be a key prediction of a higher-derivative theory of gravity
recently proposed by a subset of the current authors \cite{BMS}.}
Here, we will show that in the model proposed in
\cite{BMS}, the conditions required for the structure formation
mechanism of \cite{NBV} to work are naturally satisfied.

Obtaining a bouncing cosmology without introducing any
pathologies is a challenging task. It is believed that superstring
theory will lead to a resolution of this problem by modifying
the ultraviolet behavior of the Einstein-Hilbert action.  Any
generally-covariant modification of gravity involves higher derivative
extensions of the Einstein-Hilbert action. Thus it is no surprise that
string theory indicates such corrections \cite{Scor,GBgrav}. However,
higher derivative theories usually contain ghosts, thereby making them inconsistent. There are two different mechanisms known which can
avoid this problem. The first method exploits the fact that gravity is
a gauge theory and therefore, even if there are ghosts, they may be
benign \cite{benign}.  It then becomes possible to avoid the
``dangerous ghosts'' by choosing special combinations of the higher
derivative corrections. Gauss-Bonnet gravity \cite{GBgrav} is a
familiar example of this class (but see also \cite{calcagni} for a more critical analysis). A second method  relies on non-perturbative 
physics and is equally applicable to gauge theories (like gravity) or non-gauge
theories, such as scalar field theories. The way non-perturbative modifications
avoid the problem of ghosts is by ensuring that the corrections do not 
introduce any new states, ghosts or otherwise. (Perturbative corrections, 
\ie up to some finite order in higher derivatives,  inevitably introduce new 
states.) A well known example of such a theory is the p-adic string theory 
where the kinetic part of the action contains an infinite series of higher 
derivative terms \cite{padic} (see also  \cite{marc,warren}). 

A non-perturbative ghost-free model of gravity containing ``string-inspired''
higher derivative corrections was recently proposed in
\cite{BMS}, it was also shown that  such theories offer a possible resolution to the big
bang singularity as they admit bouncing solutions.  
The discussion of \cite{BMS} was in the context of radiative
matter. However, typically
the energy density involved during the bounce is of the order of the
cut-off scale which could be of the order of the string scale. It is
then natural to imagine that for a string-scale energy density one would
excite at least the fundamental strings. If we describe the bounce in
terms of three spatial dimensions, assuming that the other spatial
dimensions are stabilized, then one could naively expect a gas of
closed strings in a Hagedorn phase during the bouncing phase. 
Here, we first generalize the analysis of \cite{BMS} and {\it show that non-perturbative gravity actions admit bouncing solutions and therefore can resolve the big bang singularity for  general forms of matter (not just radiation),  a stringy Hagedorn phase for instance.} Then, we
show how the structure formation scenario of \cite{NBV} can be
implemented. 

As we shall argue, to generate the appropriate metric fluctuations, we would require  the bounce to  occur just after the onset of the Hagedorn phase, when  although the fluctuations in  energy is dominated by closed string excitations, the energy density itself is still  dominated  by the massless string modes (radiation)\footnote{Briefly (see appendix \ref{transition}), the reason this happens is because the energy fluctuations are proportional to the  heat capacity which in the 
Hagedorn phase goes as $T_H/(T_H-T)$, while the energy only goes as a logarithm of this ratio. Therefore, 
as the temperature reaches close to the limiting Hagedorn temperature the 
heat capacity corresponding to the excitations of the massive modes catches 
up with those of the massless modes much faster than the energy itself.}. There are three
reasons. First, in order to match the current observations, the amplitude of
the metric fluctuations, which will be determined by the ambient
temperature, must be small. It turns out that this is easier to achieve
if the bounce occurs close to the onset of the Hagedorn phase.
Secondly, if the energy density is dominated by the Hagedorn gas of closed strings,
then it is hard to keep the dilaton fixed, because in the Einstein
frame strings couple to the dilaton and will therefore source it. 
Finally, in order to apply local physics we need to define a local
density which can only be done if energy goes as volume, as it does for
radiative matter. The crucial point to note is that all the conditions required for the success of the model are met near the onset of the Hagedorn phase.

The paper is organized as follows: In section \ref{background} we first review non-perturbative gravity models and then discuss the background bouncing geometries that emerges in these models. In section \ref{hagedorn} we discuss the spectrum of stringy perturbations that can be generated in the bouncing phase if the bounce occurs within the Hagedorn phase. In particular we estimate the spectral tilt and amplitude for some specific cases. We  conclude in section \ref{conclusion} by summarizing the new structure formation mechanism based on singularity free bouncing universe scenario, and also point out possible caveats in the model. There are three appendices devoted to discussing, the technical details of obtaining the bouncing solutions (appendix \ref{approximate}),  the thermodynamics near the onset of the Hagedorn phase (appendix \ref{transition}), and the dynamics of possible light moduli fields during the course of the bounce (appendix \ref{moduli}).

\section{Non-Perturbative Gravity and Bouncing Universes}\label{background}
\subsection{Review}
In string theory, higher-derivative corrections to the Einstein-Hilbert
action appear already classically (\ie, at the tree level), but we do
not preclude theories where such corrections (or strings themselves)
appear at the loop level or even non perturbatively.  From
string field theory~\cite{sft} (either light-cone or covariant) the
form of the higher-derivative modifications can be seen to be Gaussian,
\ie there are $e^\Box$ factors appearing in all vertices (e.g.,
$(e^\Box\phi)^3$). These modifications can be moved to kinetic terms by field
redefinitions ($\phi\to e^{-\Box}\phi$). The nonperturbative gravity
actions that we consider here will be inspired by such  stringy kinetic terms~\cite{BMS}. In this context we note that similar non-local modifications of gravity has been considered previously in the literature \cite{expo}, while only recently possible modifications to the evolution during the Hagedorn phase due to (perturbative) higher derivative corrections was considered in \cite{borunda}.  

It was also noticed in Ref~\cite{BMS} that if we wish to have both a
ghost free and an asymptotically free theory of gravity\footnote{While perturbative unitarity requires the theory to be ghost free, in order to be able to address the  singularity problem in General Relativity, it may be desirable to make gravity weak at short distances, perhaps even asymptotically free \cite{weinberg}. }, one has little choice
but to look into gravity actions that are non-polynomial in
derivatives, such as the ones suggested by string theory.  Thus we start by considering the simplest non-polynomial
generalization of the Einstein-Hilbert action
\be 
S=\int d^4x\ \sqrt{-g}F(R)\,, 
\label{action}
\ee 
where
\be
F(R)=R+\sum_{n=0}^{\infty}{c_{n}\over M_{\ast}^{2n}}R\Box^{n}R\,,
\label{non-pert}
\ee
and $M_{\ast}$ is the scale at which non-perturbative physics
becomes important. $c_n$'s are typically assumed to be $\sim {\cal O}(1)$
coefficients.  It is convenient to define a function,
\be
\Ga(\la^2) \equiv \left(1-6\sum_0^{\infty}c_i\LF{\la\over M_{\ast}}
\RF^{2(i+1)}\right)\,.
\label{final-eqn}
\ee
One can roughly think of $p^2\Ga(-p^2)$ as  the modified inverse propagator for gravity (see \cite{BMS} for details). 
In \cite{BMS} it was shown that if $\Ga(\la^2)$ does not have any
zeroes, then the action is ghost-free\footnote{One may also worry about classical instabilities which usually plague higher derivative theories, generically they go by the name of Ostrogradski instabilities (see \cite{ostro} for a review). However, these are basically the classical manifestations  of having ghosts in your theory (non-unitarity or having negative norm states can be exchanged for having arbitrarily large negative energy states, which classically manifests as a catastrophic instability). Since the non-local theories under consideration do not contain any ghosts we also do not expect to find such instabilities. One way to see how these theories may avoid the Ostrogradski instability argument, which is valid for finite higher derivative theories, is by noting that one cannot construct the usual Ostrogradski Hamiltonian because one cannot identify a  ``highest derivative'' in such non-local actions. Also, in arriving at the  Ostrogradski Hamiltonian, one assumes that  all the  derivatives of the field (except the maximal one) are independent canonical coordinates. This is no longer true for theories with infinite derivatives. For instance,  one cannot independently choose an initial condition where one can specify arbitrarily  the values of all the derivatives of the field. See \cite{padic} for a nice discussion regarding this.}, and in fact one introduces no other states other than the usual massless graviton (there are no extra poles in the propagator). One can also check that if $\Ga$ goes like an exponential, as expected from stringy arguements, then not only is it ghost-free, but it also describes and asymptotically free theory. 

Since we are interested in cosmological solutions, and in particular
homogeneous and isotropic cosmologies with a metric (for simplicity we
consider the case of a spatially flat universe) given by
\be
ds^2 \, = \, - dt^2 + a(t)^2 \bigl(dx^2 + dy^2 + dz^2 \bigr) \, ,
\ee
where $a(t)$ is the scale factor, it is sufficient to look at the
analogue of the Hubble equation for the modified action
(\ref{action},\ref{non-pert})
\footnote{Just as in  ordinary Einstein gravity, here also the Bianchi identities (conservation equation) ensure that for FRW metrics the $\ti{G}_{ii}$ equation is automatically satisfied when the $\ti{G}_{00}$ equation is.}. 
This equation is \cite{schmidt,BMS}
\be
\ti{G}_{00} \, = \, F_0R_{00}+{F\over 2}-F_{0;\ 00}-\Box F_0-
\2\sum_{n=1}^{\infty}F_n\Box^n R-{3\over 2}
\sum_{n=1}^{\infty}\dot{F}_n\dot{(\Box^{n-1}R)} \, = \, T_{00}\,,
\label{tildeG00}
\ee
where we have defined
\be
F_m \, \equiv \, \sum_{n=m}^{\infty}\Box^{n-m}{\p F\over \p \Box^n R}\,.
\ee
It was shown in \cite{BMS} that (\ref{tildeG00}) admits exact bouncing
solutions of the form
\be
a(t) \, = \, \cosh\LF{\la t\over \sqrt{2}}\RF \,.
\label{cosh}
\ee 
in the presence of radiative matter sources and a non-zero
cosmological constant. Note that this is an exact
result, but other approximate bouncing solutions can now also be constructed in the presence of arbitrary matter sources, which can match the gradual contraction before and expansion after the bounce. This is important for several reasons. Firstly, it shows that the cosmological constant is not necessary for the bounce as was  already argued in  \cite{BMS}. Secondly, it shows that one can have bounce not only in the presence of radiation but also other sources of matter, such as a stringy hagedorn phase which is expected in the early universe. Finally, we can obtain bouncing solutions for a much larger class of non-perturbative actions thereby exhibiting the intrinsic quality of these corrections to resolve singularities.  We now discuss these approximate solutions. 

\subsection{Approximate Bouncing Solutions for Generic Matter Sources:Energy and Distance Scales }

For the purpose of this paper, since we are interested
mainly in the bouncing phase, we will assume the bounce to be of the form
(\ref{cosh}) without loss of any generality. The crucial point is to note that the entire dynamics can be analyzed by 
separating the time evolution into
three different regimes. Very close to the bounce all
the non-perturbative terms become important and one can approximately
compute both the left and the right hand side of the modified Hubble
equation (\ref{tildeG00}) to obtain solutions which are valid close to
the bounce, i.e. for $\la |t|<1$. However, it is easy to check (see
Appendix A for details) that the non-perturbative corrections fall
faster than the Einstein tensor as one goes away from the bounce, and
that around $\la |t|\sim 1$, Einstein gravity takes over. Thus for
$\la |t|>1$, both in the contracting and expanding phase, we have the
usual evolution governed by ordinary gravity coupled to matter. The main task therefore is to understand the evolution near the bounce (if there exists one).  

For this purpose, we will not assume any specific form for the energy density, namely
the source term, see the right hand side of (\ref{tildeG00}), but just
that it is a function of the scale factor:
\be
T_{00} \, = \, \rho(a) \, = \, \rho(a(t)) \, .
\ee 
What we now can do is to expand both the left and right hand side of (\ref{tildeG00}), \ie, $\tilde{G}_{00}$ and $ \rho(a(t))$ as a power series in $t^2$ for the bouncing ansatz (\ref{cosh}). Then by matching the coefficients of the first few powers in $t^2$, we can determine the dynamical parameters (see appendix \ref{approximate} for details). In principle this calculation can be done for any matter density, as long as we know $\rho(a)$, but for the cases of practical interest $\rho(a)$  behaves as an ideal gas with a specific equation of state parameter $\om$
\be
p \, = \, \om\rho   \, ,
\label{state}
\ee
where $p$ is the pressure of the fluid.  This implies
\be
\rho \, = \, \rho_ba^{-3(1+\om)}
\ee
where $\rho_b$ is the maximal density attained at the bounce point. Assuming such an ideal gas approximation,  after some algebraic manipulations   one finds that the time scale of the bounce, $\la^{-1}$, is determined by an equation of the form 
\be
{\cal K}\LF{\la^2\over M_{\ast}^2}\RF \, = 0
\label{lambda-sol}
\ee
where ${\cal  K}$ is a dimensionless function determined entirely in terms of $\Ga$ (see appendix \ref{approximate}) \ie, the non-perturbative corrections. This is not surprising as it is these corrections which are responsible for causing the bounce in the first place. Also, one straight away sees that since the coefficients in the function ${\cal  K}$ are typically expected to be $\sim \cO(1)$, we should have
\be
\la \, \sim \, M_{\ast} \, .
\label{omega-M}
\ee
unless there is some fine-tuning involved. Next, we find that the maximal energy density, $\rho_b$ , is given by an equation of the form
\be
\rho_b \, = \, {3M_p^2\la^2{\cal G}({\la^2\over M_{\ast}^2})\over 8} >0\, .
\label{rho-sol}
\ee
where ${\cal G}$ is again a dimensionless function which can be expressed in terms of $\Ga$, see appendix \ref{approximate} for details. It's specific expression is complicated and not very illuminating, but again one does not need to know the details to understand the basic physics. For instance what is important is to be able to determine the scale of energy density at the bounce point. Again, if no fine-tuning is involved, then ${\cal G}\sim \cO(1)$ and we have
\be
\rho \, \sim \, M_p^2\la^2 \, \sim \, M_p^2M_{\ast}^2 \, .
\label{rho-app}
\ee

It is now clear when we can start  to see some effects of a 
stringy Hagedorn phase: During the phase of radiation domination, the 
density is given by $\rho\sim T^4$. Therefore we would expect stringy 
effects to show up if $\rho_b> T_H^4\sim M_s^4$, or in other words as long as
\be
M_{\ast} \, > \, {M_s^2\over M_p} \, .
\label{scales}
\ee
If the higher derivative corrections  are genuinely stringy in origin, 
then $M_{\ast} \sim M_s$, and one sees that (\ref{scales}) is easily satisfied
as the string scale is expected to be less than the Planck scale. Thus, 
in general, one would indeed expect our bouncing universe to probe the 
very stringy Hagedorn phase of matter.

However, as we will see later, for phenomenological reasons we will  
be interested in the regime where the bounce probes just the onset of the 
Hagedorn phase. This is possible if one allows for some fine-tuning in either (\ref{lambda-sol}) or (\ref{rho-sol}). For instance if $\la\ll M_{\ast} \sim M_s$, then one can easily have $\rho_b\sim T_H^4\sim M_s^4$ as the last ``$\sim$'' in (\ref{rho-app}) is no longer true. 

Finally, we find that in order to have a  consistent bouncing solution we also have a constraint   equation\footnote{This is not surprising, as such constraints were found  when
 dealing with the exact solutions in \cite{BMS}, and in fact coincide 
for these special cases.} which looks like
\be
1+\om \, = \, {16{\cal F}({\la^2\over M_{\ast}^2})\over 3{\cal G}({\la^2\over M_{\ast}^2})}>0
\label{omega-sol}
\ee
(${\cal F}$ is again a dimensionless function expressible in terms of $\Ga$). For a given equation of state, together with (\ref{lambda-sol}) it over-determine $\la^2$, thereby restricting the set of non-perturbative actions ($\Ga$'s) which can exhibit bouncing solutions of the nature (\ref{cosh}), for a given equation of state.  However firstly, this shows that non-perturbative actions can resolve the singularity not only in the case of radiation (plus a cosmological constant), but also for other general matter sources.  In particular one can check that for a large number of ghost free actions one can find solutions with $\om=1/4$, that is for pure radiation, while we can also find actions admitting solutions deep inside a stringy Hagedorn phase where energy remains approximately constant, \ie $\om=1/3$.

Secondly, since we have the freedom to choose $\om$, even while restricting ourselves  to the dominant energy condition (the inequalities in (\ref{omega-sol}, \ref{rho-sol}) which essentially implies that for the $\la$ which solves the equations, ${\cal F,G}$  have to be   positive), 
 we can find a much larger class of actions which manifest the 
bouncing behaviour, at least for some $\om$.  This  illustrates the intrinsic capability of non-perturbative higher derivative corrections to resolve the big bang 
singularity problem. 
We note, in passing, that for usual Einstein gravity or $R^2$ gravity, 
the dominant energy condition cannot be satisfied. Indeed, one does not have a non-singular bounce in those theories.

Before concluding this section, let us briefly discuss about another important physical scale, that of  the size of the universe, $R_b$, at
the bounce point. In the context of a bouncing universe, it is
quite natural to have this scale much larger than the string scale.
In order for the structure formation mechanism of \cite{NBV} to work,
we require that $R_b$ is larger than $1$mm. In the context of
the string gas cosmology setup proposed in \cite{BV,NBV}, this
is not natural. Obtaining such a large initial universe in the
context of string gas cosmology leads to a similar entropy problem
as is faced in Standard Cosmology (for a possible solution of this
problem making use of a gas of branes, see \cite{Natalia}). An interesting
aspect of our bouncing universe proposal is that there is no entropy
problem, as long as the universe in the contracting phase started out
large (which, in the context of a cold initial state, is quite
natural).

One more comment: so far we have not discussed the role of the
dilaton $\phi$. We have tacitly assumed that the dilaton has been fixed, 
such that the string coupling  $g_{s}\sim e^{\phi/M_{P}}$ is constant.
In fact, we can allow the dilaton to be dynamical
near the bounce, see Appendix \ref{moduli}; typically the dilaton attains a constant value before and after the bounce and jumps by a finite amount during the course of the bounce. The amount of jump however depends on the initial conditions and can be made to be very small, so that it becomes a good approximation to treat the dilaton profile near the bounce to be a constant.

\section{Hagedorn Phase and Structure formation}\label{hagedorn}
\subsection{The New Cosmological Scenario}
The bouncing solution obtained in the previous section has interesting
consequences for cosmology if the energy density at the bounce point
is large enough to lead to a Hagedorn phase of string matter. 

We begin with a sketch of the space-time diagram obtained in our 
scenario (Fig. 1). The vertical axis in Fig. 1 denotes physical time, 
with the origin of time chosen such that $t = 0$ is the bounce point. 
The horizontal axis denotes comoving spatial coordinates. The two dotted
vertical lines labeled by $k_1$ and $k_2$ denote the comoving wavelengths
of cosmological fluctuations with comoving wavenumber $k_1$ and $k_2$.
The solid curve is the Hubble radius, defined as $\cal{H}^{-1}$,
where $\cal{H}$ is the Hubble expansion rate in conformal time. The key
point is that the Hubble radius becomes infinite at the bounce point.
In particular, fluctuations on all scales of interest in cosmology today
are sub-Hubble in a time interval about the bounce point. The
evolution of the Hubble radius as a function of time for $t > 0$ is
analogous as in the string gas cosmology setup discussed in \cite{NBV}, 
where it was shown that string thermodynamic fluctuations during an early
quasi-static Hagedorn phase can lead to a scale-invariant spectrum of
curvature fluctuations.
\begin{center}
\begin{figure}
\includegraphics[height=6cm]{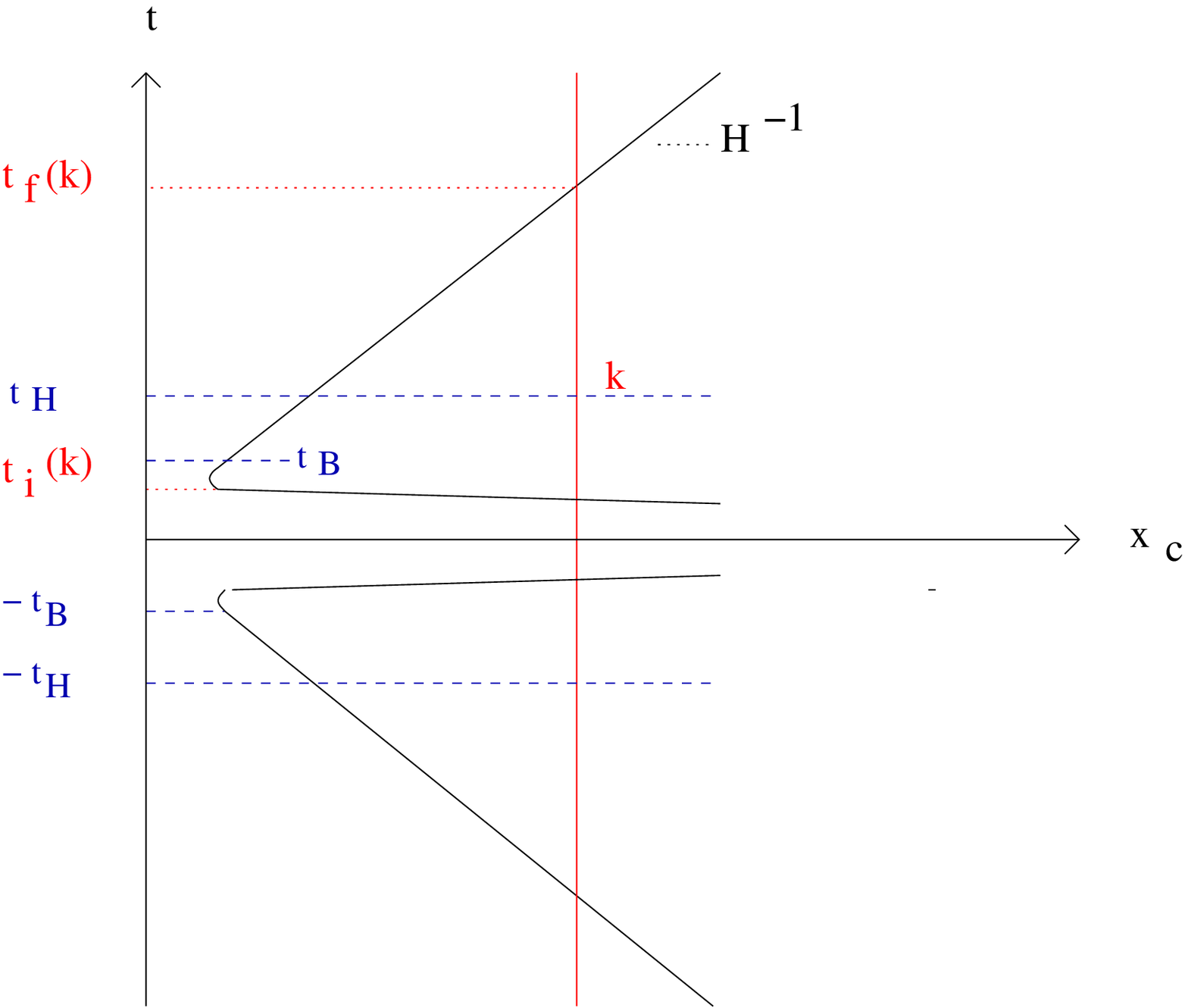}
\begin{caption}
{\small Space-time diagram (sketch) showing the evolution
of a fixed comoving scale in our bouncing cosmology under the assumption
that the bounce phase lies entirely within the Hagedorn phase. 
The vertical axis is time, the horizontal
axis is comoving distance. The bounce time is chosen to
be $t = 0$. The Hagedorn phase corresponds to
the time interval $-t_H < t < t_H$, the bounce phase is the
interval $-t_B < t < t_B$. Outside of the bounce phase, the
Hubble radius evolves as in a radiation phase of standard
cosmology The black solid curve represents the Hubble radius $H^{-1}$.
The red vertical line labelled by $k$ depicts a fixed comoving scale.
This scale is smaller than the Hubble radius in the
central region of the bounce, exits the Hubble radius at the
time $t_i(k)$, and re-enters at a late time $t_f(k)$.}
\end{caption}
\label{fig:1}
\end{figure}
\end{center}
Let us recall the origin of the Hagedorn phase \cite{BV}: In the context of
perturbative string theory, there is a maximal temperature $T_H$, the
Hagedorn temperature \cite{Hagedorn}, for a string gas in thermal 
equilibrium. As the universe contracts in our scenario, the temperature
initially increases as in the usual radiation phase of standard cosmology.
The energy is in the radiative degrees of freedom, the momentum modes of
the strings. As the temperature approaches $T_H$, it becomes possible
to excite the string oscillatory and winding mode degrees of freedom.
In fact, close to the Hagedorn temperature, most of the energy in the
string gas is contained in winding strings \cite{Deo,MT}.

The crucial point of the mechanism of \cite{NBV} is the following: provided
that our three large spatial dimensions are compact,  the heat capacity
$C_V$ of a gas of strings in thermal equilibrium scales as $r^2$ with the
radius of the box. The heat capacity determines the root mean
square  mass fluctuations via
\be \label{massfluct}
\langle\bigl( \delta M \bigr)^2\rangle \, = \, T^2 C_V \, , 
\ee
which then in turn induce metric perturbations. In longitudinal
gauge (see e.g. \cite{MFB,RHBrev} for reviews of the theory of cosmological 
perturbations), and in the absence of anisotropic matter stress at late
times, the metric takes the form
\be
ds^2 \, = \, - (1 + 2 \Phi) dt^2 + a(t)^2 (1 - 2 \Phi) d{\bf x}^2 \, ,
\ee
where $t$ is physical time, ${\bf x}$ are the comoving spatial
coordinates of the three large spatial dimensions, $a(t)$ is the
cosmological scale factor and $\Phi({\bf x}, t)$ represents the
fluctuation mode. On scales smaller than the Hubble radius,
the metric fluctuations are driven by the matter fluctuations,
on super-Hubble scales it is the metric fluctuations which are
dominant since the matter oscillations have frozen out (thermal fluctuations cannot keep up with the Hubble expansion). Thus,
to compute the late-time spectrum of metric fluctuations,
the philosophy adopted in \cite{NBV} (see  also \cite{RB} for
a corresponding treatment of fluctuations in inflationary cosmology and \cite{pogosian} for a discussion on the spectrum of thermal fluctuations corresponding to ordinary particles)
is, for inhomogeneities on a fixed comoving scale, to follow the 
matter fluctuations until the scale exits the Hubble radius at time
$t_i(k)$, and to use the Einstein constraint equations
to determine the metric fluctuations at that time. The metric fluctuations
will then be conserved on super-Hubble scales. 

As long as the background dilaton velocity is negligible,
the Einstein constraint equations (specifically, the relativistic
Poisson equation) 
\be \label{constr}
\nabla^2 \Phi \, = \, 4 \pi G \delta \rho \, , 
\ee
(where $\rho$ is the energy density)
then imply that the power spectrum of the metric fluctuation
variable $\Phi$ is scale-invariant (for details on how to
get from the spectrum of mass fluctuations to the power
spectrum of $\Phi$ see \cite{BNPV2}).

In order for (\ref{constr}) to apply, it is important that metric
and matter fluctuations are related via the Einstein equations, and
not e.g. via the dilaton gravity equations \cite{TV} (see \cite{Betal}
for a detailed discussion of this point). In our bouncing cosmology,
the dilaton is taken to be constant throughout the evolution of the universe.
Thus, there are no dilaton corrections to (\ref{constr}). We also note that the higher derivative corrections to (\ref{constr}) are suppressed by $(\n^2/M_{\ast}^2)\sim (k/M_{\ast})^2$ (see \cite{BMS} for a quantitative discussion). The comoving  scales that we are interested in corresponds to the Hubble radius today, and one can easily check that this ratio is hugely suppressed (see section \ref{transition}  for instance) in a non-inflationary scenario such as ours. In addition,
the fact that there are no additional dynamical degrees of freedom compared
to those in Einstein's theory ensures that we  need not worry about extra unstable (ghost like)  or spurious  excitations of the metric. 

We are thus led to the following structure formation scenario: The universe
begins in a cold contracting phase. The initial Hubble radius is set
by the initial gas density, and it is natural to assume that the size
of space is at least as large as the initial Hubble radius. If we begin
the evolution at a temperature lower than the present radiation temperature,
we are then ensured that the size of space at the bounce is sufficiently large
to contain all fluctuation modes of current interest in cosmology
\footnote{Note, in particular, that the bounce radius $R_b$ is
{\it not} set by the physics which determines the bounce, but
is in fact much larger than the length scale of the
local bounce physics.}. Local
processes in at the beginning or during the course of the
contraction phase establish thermal equilibrium over a scale which
contains all fluctuation modes of interest to us. Matter remains homogeneous
and isotropic throughout the bounce. In particular, during the
Hagedorn phase of the bounce, we have a gas of strings. In the same
way that a network of strings forming at a phase transition in
a field theory model with cosmic strings \cite{CSreviews} will contain
strings of length larger than the Hubble radius (in particular, it
will contain winding modes), after the transition to the Hagedorn
phase, our gas of fundamental strings will contain winding modes, and
it is the presence of these winding modes which are the key
ingredient to obtain the form of the specific heat we are using. Thus, 
we are able to use the results of string thermodynamics to compute the
spectrum of matter perturbations which then seed the scalar metric
fluctuations.

\subsection{Thermodynamics, a Brief Review}

In order to compute the power spectrum of density fluctuations we have to 
compute the fluctuations in string energy, 
$\De E$, for these closed string modes inside an arbitrary volume $V$. Once we obtain $\De E(V)$, there 
is  a well defined prescription on how to go over to the power spectrum ${\cal P}(k)$ as described in the 
next subsection. In order to obtain $\De E$ and eventually the power spectrum 
we have to understand the thermodynamics of the Hagedorn phase and here we 
start with a brief review of the same. 
  
Specifically we are interested in a situation 
where we have $d=3$ large but compact spatial dimensions. The entropy $S$ of 
a string gas in thermal equilibrium in this background was derived in 
\cite{Deo2} (also see \cite{NBV,Ali}). During the Hagedorn phase,
and in the case of a volume which is large in string units, the
result is 
\be
S \, = \, {\cal E} + bv -
{\cE^{2d-1}\over (2d-1)!}\exp\LT-{c(\cE-kv)\over v^{2/3}}\RT \, ,
\label{entropy}
\ee
where we have introduced the dimensionless variables
\be
\cE \, \equiv \, {E\over T_H} \mx{ and } v \, \equiv \, VT_H^d \, ,
\ee
for convenience. $V$ is the volume of the $d$ large dimensions, and 
$b,c,k$ are ${\cal O}(1)$ constants (see \cite{Deo,NBV,Ali}).  This formula 
is valid in the regime $\cE > kv$ (which is characteristic of
the Hagedorn phase) and $v \gg v^{2/3}$ (which means that the volume of
the universe is much larger than the string length). 

Once we have the microcanonical entropy $S(E,V)$  as a function of the 
energy and the volume, it is straight forward to compute all the 
thermodynamic quantities. In particular we find that the inverse
temperature $\beta$ is given by
\be
\bb \, = \, {\p S\over \p E}
\ee
which implies
\be  
{\bb\over \bb_H} \, = \, 1 + 
{c\cE^{2d-1}\over (2d-1)! v^{2/3}}\exp\LT-{c(\cE-kv)\over v^{2/3}}\RT \, ,
\label{temperature}
\ee
where we have  used $v\ll \cE$. Using the same approximation we find from 
(\ref{temperature}) that the specific heat (or rather the heat capacity) 
is given by
\be
C_v \, \equiv \, \bb_H\LF{\p \cE\over \p T}\RF_v
\, = \, {(2d-1)!v^{2/3}\over cT(T_H-T)} \, ,
\label{heat-capacity}
\ee
which gives the characteristic scaling $C_V \sim r^2$ as a function of the
radius of the volume which, by the discussion in the previous
subsection implies a scale-invariant spectrum of metric fluctuations.

If the decrease in the Hubble radius at the end of the Hagedorn phase
is not instantaneous, then the time $t_i(k)$ (and thus also the 
corresponding temperature $T(t_i(k))$ will depend on $k$. This will
lead to a red tilt in the spectrum. In section (\ref{tilt}) 
we will see that using (\ref{temperature}) 
and (\ref{heat-capacity}) one can calculate the spectral tilt in the 
bouncing universe scenario quite precisely.

\subsection{Spectrum of Perturbations}

The dimensionless power spectrum of scalar metric fluctuations is defined by
\be \label{spectrum}
P_{\Phi}(k) \, \equiv \, k^3 |\Phi(k)|^2 \,
\ee
where $\Phi(k)$ is the k'th Fourier mode of
$\Phi(x)$, and we are using the convention of defining the Fourier
transform of a function $f(x)$ including a factor of the
square root of the volume of space. 

Making use of the relativistic Poisson equation (\ref{constr}), 
(\ref{spectrum}) becomes
\be \label{power1}
P_{\Phi}(k) \, = \, 16 \pi^2 G^2 k^{-1}
|\delta \rho(k)|^2 \, .
\ee
The Fourier space density perturbation of wavenumber $k$ can be determined from
the mean square mass fluctuation $|\delta M|^2(r)$ in a sphere of
radius $r(k) = k^{-1}$ via
\be
|\delta \rho(k)|^2 \, = \, k^{3}|\delta M|^2(r(k))  \, ,
\ee
and hence (\ref{spectrum}) becomes
\be \label{power2}
P_{\Phi}(k) \, = \, 16 \pi^2 G^2 k^2 (\delta M)^2(r(k)) \, .
\ee
Plugging the expression (\ref{heat-capacity}) for the specific heat capacity of
a string gas into (\ref{massfluct}) and then into (\ref{power2}) we finally
obtain the following result for the power spectrum \cite{NBV,BNPV2}:
\be \label{power}
P_{\Phi}(k) \, = \, 1920 \pi^2 c^{-1} G^2 T_H^4 \bigl( k r(k) \bigr)^2
{T \over {T_H (1 - T/T_H)}} \, ,
\ee
where for each value of $k$, the temperature $T$ is to be evaluated
at the time that the mode $k$ exits the Hubble radius at time $t_i(k)$.

We see immediately that as a consequence of the specific stringy scaling
of the heat capacity, we obtain a scale-invariant spectrum of
fluctuations in the approximation when $T(t_i(k))$ is independent of $k$
as it will be if the Hubble radius drops abruptly. If the decrease of
the scale factor is not instantaneous, then a small red tilt of the
spectrum will be induced since smaller length scales exit the Hubble
radius later, and the temperature then is smaller. We will come back
to the issue of the magnitude of the spectral tilt shortly. 

The overall amplitude of the spectrum is, as in \cite{NBV,BNPV2}, determined
by two factors, firstly the factor $G^2 T_H^2$ which is the fourth power
of the ratio of the Planck length and the string length, and secondly
the final factor in (\ref{power}) which depends on how close the
temperature at the bounce point is to the Hagedorn temperature.

\subsection{Spectral Tilt: General Formula} \label{tilt}

In this subsection we compute the spectral tilt $\eta_s - 1$ which is
defined by
\be
P(k) \, \sim \, k^{\eta_s - 1} \, .
\ee
To keep things general we do not specify the function $T=T(a)$. 
The spectral tilt can be calculated straightforwardly from (\ref{power}).
The key factor is the factor $1 - T/T_H$ in the denominator of the
expression. This is the term that carries the dominant dependence on $k$.
The reason is that at all times $t_i(k)$, the temperature is very
close to the Hagedorn temperature so that the only other factor which
depends on $k$, namely the factor $T$ in the numerator of (\ref{power})
can be taken to be constant. From (\ref{power})
\be 
\ln {\cal P} \, = \, \ln C - \ln(1-T/T_H) \, ,
\ee
where the constant $C$ stands for the product of all terms in (\ref{power})
which do not depend on $k$. Hence,
\be
{d\ln {\cal P}\over dk} \, = \, {d\ln {\cal P}\over da}{da\over dk} \, ,
\ee
where the comoving wave-vector $k$ exits the Hubble radius when the scale 
factor $a$ is given by 
\be
k \, = \, Ha \, = \, \dot{a}
\label{crossing}
\ee
which implies
\be
\dot{k} \, = \, \ddot{a} \, .
\ee
Note that (\ref{crossing}) determines $k$ as a function of the scale factor 
$a$, and through its dependence on time, also as a function of time, 
$k=k(a)=k(a(t))$.

From (\ref{crossing}) we find 
\be {da\over dk} \, = \, \left({da\over dt}\right)/{dk\over dt} \,
= \, {\dot{a}\over \ddot{a}} \, .
\ee
Thus,
\be
{d\ln {\cal P}\over dk} \, = \, {d\ln {\cal P}\over da}{\dot{a}\over \ddot{a}} 
\, .
\ee
Now, straightforwardly one finds
\be
{d\ln {\cal P}\over da} \, = \, {T'T_H\over (T_H-T)T} \, ,
\ee
where $'$ denotes the differentiation with respect to the scale factor. 
Putting all the things together we have
\be
{d\ln {\cal P}\over dk} \, = \, {\dot{a}\over \ddot{a}}{T'T_H\over (T_H-T)T}
\, ,
\ee
and, finally,
\be
\eta_s-1 \, \equiv \, {d\ln {\cal P}\over d\ln k} \, = \, 
k{d\ln {\cal P}\over dk} \, = \, {\dot{a}^2\over \ddot{a}}{T'T_H\over (T_H-T)T}
\label{spectral}
\ee
This is a general result which holds for any $T(a)$ and $a(t)$. One  can 
further simplify (\ref{spectral}) using the following redefinition:
\be
a(t) \, = \, e^{\cN(t)}
\ee
One can check that the expression for the spectral tilt then reduces to
\be
\eta_s-1 \, = \, \left[{H^2\over H^2 + \dot{H}}\right]
\left[{1 \over T}{dT \over d\cN}{T_H\over (T_H-T)}\right] \, \equiv \, XY \, ,
\label{bounce-spectral}
\ee
where $X (Y)$ denotes the term in the first (second) set of brackets.
Note that while the first term depends on how the scale factor evolves 
with time, the second term primarily depends on how the temperature 
changes with the expansion of the universe, $T(\cN)$. 
\subsection{Spectral Tilt and Amplitude during Bounce}~\label{STB}

In this section we evaluate the magnitude of the spectral tilt  in our
scenario keeping in mind the various consistency requirements of the mechanism and the fact that we have to reproduce the correct amplitude of fluctuation in CMB. We recall that, as in inflationary cosmology, the cosmological
fluctuations on a fixed comoving scale $k$ freeze out at the time $t_i(k)$ 
when their wavelength crosses the Hubble radius. This is similar to what
happens in the case of cosmological inflation. However, whereas in inflationary
cosmology physical scales are increasing exponentially while the
Hubble radius stays approximately constant, and thus scales are ``expelled'' 
from the Hubble volume, in a bouncing scenario such as the one presented
in this paper, scales exit the Hubble
radius in spite of the fact that during the relevant time interval
the universe (and hence the physical scales) expands very little. Instead, 
the Hubble radius sharply decreases to a microscopic value starting from 
infinity (at the bounce point).  

Since the scale of the bounce is governed by $\la\sim M_{\ast}$ which 
is expected 
to be of the order of the string scale, approximately in a string time  
the Hubble radius decreases to string length from infinity.  All the scales 
which  we observe today exit the Hubble radius in the process 
\footnote{even if $\la$ is smaller than $M_{\ast}$ as required in our 
scenario, the picture remains true.}.

Let us first compute the temperature dependent factor $Y$ in the equation
(\ref{bounce-spectral}) for the spectral tilt. As we will soon see phenomenological constraints will tell us that the bounce has to occur just after the onset of the Hagedorn phase when  the background evolution is still governed by radiation (massless modes) as explained in the appendix \ref{transition}.  The task of computing $Y$ is then easy as the temperature is just given by
\be
T \, = \, T_b e^{-\cN}\approx T_He^{-\cN} \, .
\ee
Hence,
\be
Y \, = \, {T_H\over T_H-T} \, \approx \, {T_H\over T_H-T_b} \, ,
\ee
since all the relevant modes are effectively expelled out of the 
Hubble radius very near to the bounce point.

For (\ref{cosh}) one can easily compute the first term, $X$, in the 
expression of the spectral tilt (\ref{bounce-spectral}), obtaining
\be
X \, \equiv \, {H^2\over H^2+ \dot{H}} \, = \, 
{\tanh^2(\la t/\stwo)\over\tanh^2(\la t/\stwo) +
\mt{sech}^2(\la t/\stwo)}\approx {\la^2 t^2\over 2 + {\cal O}(\la^4 t^4)} \, .
\ee
Putting everything together, we have a rather simple expression for the 
spectral tilt evaluated at a length scale given by the comoving wavenumber
$k$ (note that the tilt in general is scale-dependent)
\be
\eta_s-1 \, = \, -{\la^2 t_k^2\over 2}{T_H\over T_H-T_b} \, \equiv \,
-{\la^2 t_k^2\over 2}{T_H\over \De T}
\ee
and we recall that the physical wavelength of the mode exits the Hubble radius 
at a time denoted by $t_i(k)$. 

We will soon find out that for the physical scales that we observe in the CMB, 
$t_i(k) << \la^{-1}$. In this approximation $t_i(k)$ is determined by:
\be
k \, = \, Ha \, = \, \dot{a} \, = \,
{\la\over \stwo}\sinh{\la t\over \stwo}\approx {\la^2 t_i(k)\over 2}
\ee
so that
\be
\eta_s-1 \, \approx \, -\la^2 t_k^2{T_H\over \De T}
\, = \, -2\LF{k\over \la}\RF^2{T_H\over \De T}
\ee
The comoving wave number corresponding to the Hubble radius today is given by
\be
k_0 \, = \, H_0 a_0 \, ,
\ee
and thus we have
\be
\eta_s-1 \, = \, -2\left({H_0a_0 \over \la}\right)^2{T_H\over \De T}
\ee
Since during the bounce phase the universe expands very little, one can easily estimate $a_0\sim T_0/T_H$ using the usual scaling during  radiation era. Then it is easy to check that (see also appendix \ref{transition})
\be
|\eta_s-1|\approx 10^{-60}\left({M_s \over \la}\right)^2 {T_H\over \De T}
\ee
Thus  we observe that the ``bounce factor'' $X$ is typically very small and  thus, as long as the temperature dependent factor 
$Y$ is not too large we will reproduce a scale-invariant spectrum.

For illustration let us consider a specific example which is consistent with all the requirements, viz. (i) we get a scale invariant spectrum, (ii) amplitude of fluctuation $\sim 10^{-10}$ as required by CMB, (iii) bounce occurs near the transition, \ie $\rho_b\sim M_s^2\sim M_p^2\la^2$, and finally (iv) the fluctuations in energy are dominated by the closed strings in the Hagedorn phase. As explained in the appendix, the last constraint is met as long as\footnote{We should point out that this is not an unnatural requirement. During the Hagedorn phase the temperature for most part remains very close 
to the Hagedorn temperature. The difference between the maximal temperature 
and the Hagedorn temperature is indeed exponentially suppressed.} 
\be
{\De T\over T_H}<10^{-30}
\ee
We will consider the limiting case, when $\De T/T_H\sim 10^{-30}$. Now the amplitude of fluctuation is approximately given by
\be
\de_{\mt{CMB}}^2\sim 10^{-10}\sim \LF{M_s\over M_p}\RF^4 {T_H\over \De T}\Ra M_s\sim 10^{-10}M_p
\ee
In other words an $M_s\sim 10^{-10}M_p$, reproduces the required amplitude of perturbations observed in CMB.  Requirement (ii) now forces $\la\sim 10^{-10}M_s$ giving rise to a very small red spectral tilt
\be
|\eta_s-1|\sim 10^{-10}
\ee

This situation can of course change once we include additional corrections to the entropy relation, $S(\cE,v)$ near the transition, or if the scales involved are different. In fact, a really interesting case emerges if one considers the string scale to be around ($Tev$) which is also interesting from the point of view of particle physics phenomenology. In this case, $\la\sim 10^{-15}M_s\sim 10^{-30}M_p$, and one obtains a spectral tilt which is just becoming observationally significant!

Before ending this section let us point out why for the phenomenological success, we require the bounce to occur close to the transition. If $\De T/T_H\ll 10^{-30}$,  for instance, then in order to get the amplitude of spectrum right, one has consider an unacceptably low string scale. Typically, deep inside the Hagedorn phase $\De T/T_H$ is indeed very small, it is suppressed by an exponential of an exponential, and thus is not phenomenologically viable.
\section{Conclusions and Discussion}\label{conclusion}
In this paper we have extended the non-singular bouncing cosmology
construction obtained in \cite{BMS} to more general matter
sources, including a gas of closed strings. We have shown that this
model leads to a simple implementation of the
string gas structure formation mechanism proposed in \cite{NBV}, provided
that the energy density at the bounce is sufficiently large to lead
a Hagedorn phase of the string gas, and the three large spatial
dimensions are compact.

The higher derivative gravity model discovered in \cite{BMS} is
distinguished by being free of ghosts. No new gravitational degrees
of freedom appear compared to those of Einstein gravity. The dilaton
can be consistently taken to be frozen at all times. Thus, it is
reasonable to assume that the metric fluctuations can be described
using the equations of general relativistic perturbation theory.

In this case, all of the conditions to obtain a scale-invariant
spectrum of scalar metric fluctuations detailed in \cite{BNPV2,Betal}
are satisfied. The matter fluctuations  in the Hagedorn phase around
the bounce time, when the Hubble radius diverges, 
are characterized by a heat capacity $C_V$ which
scales as $r^2$ with the size of the region \cite{Deo2,NBV,Ali}. Via
the relativistic Poisson equation, these matter fluctuations induce
a scale-invariant spectrum of metric fluctuations which propagate to
late times. being squeezed, but without modification of the spectral shape.
The spectrum has a red tilt, like in the string gas cosmology scenario
\cite{NBV,BNPV2}. However, the magnitude of the tilt - calculated in
this paper - is very small.

Assuming that the initial state is chosen very early in the
contracting phase, when the matter density is smaller than the
current density, and assuming that the spatial size is at least one
Hubble volume at that time, we obtain a cosmology which resolves
the horizon and entropy problems of standard cosmology \footnote{By 
the entropy problem we mean the problem of obtaining a universe sufficiently
large at the current temperature to include our current Hubble
volume (see also \cite{Natalia} for another proposal to solve the
entropy problem without inflation, making use of non-trivial dynamics
of the extra spatial dimensions which string theory predicts).}.
Thermal equilibrium over large distances is established in the
contracting phase, thus allowing us to use equilibrium thermodynamics
of strings in the Hagedorn phase, even if the latter is of a duration
short compared to the physical length at the bounce time of perturbations
relevant to current cosmology. Thus, our model provides an explicit
toy model which demonstrates, in the context of an explicit action
for the background space-time, that all the criteria required to obtain
a scale-invariant spectrum in the scenario of \cite{NBV} can be
realized, and thus successfully addressed the concerns raised in
\cite{KKLM}. 

\section*{Acknowledgments} Tirtho would like to thank Natalia  (Shuhmaher) and Keshav (Dasgupta) for having lively discussions during the project. Tirtho would also like to thank Tufts University for their continued hospitality during his frequent visits to Boston. The research at McGill was supported in part by NSERC, by  McGill University, by the Canada Research Chair program, and by the  FQRNT. The research of AM is partly supported by the European
Union through Marie Curie Research and Training Network
``UNIVERSENET'' (MRTN-CT-2006-035863). WS was  supported in part by NSF Grant PHY-0354776.
\appendix
\section{Exact and Approximate Bounce Solutions}\label{approximate}

In the previous work \cite{BMS}, the presence of a cosmological constant 
was required for the late-time consistency of the ansatz (\ref{cosh}) for
the solution. To see this consider the behavior at large times. It is 
clear that the scale factor tends to a de Sitter solution:
\be
a(t) \, \ra \, {1\over 2}e^{\sqw t}\, ,
\label{inflation}
\ee
while one can check that all the higher curvature terms vanish:
\be
\ti{G}_{00}^{HD} \, \equiv \, \sum_{n=0}^{\infty}\ti{G}_{00}^n\ra 
\mt{sech}^2\left(\sqw t\right)\ra 0\, .
\label{hc-energy}
\ee
As a result, we are just left with Einstein's theory of gravity. In the 
absence of the cosmological constant,  at large times when the higher 
derivative corrections become small, we will instead make a transition to 
the usual radiation dominated era of Einstein gravity.

For the cosine hyperbolic ansatz the modified Einstein tensor looks like 
\begin{eqnarray}
\ti{G}_{00} \, &=& \, L_0+L_2\mt{sech}^2\left(\la t/\stwo\right) +
L_4\mt{sech}^4\left(\la t/\stwo\right) +
L_6\mt{sech}^6\left(\la t/\stwo\right) \\
&=& \, C_0+C_2\la^2 t^2+C_4\la^4 t^4+{\cal O}(\la^6 t^6) \, ,
\label{Gseries}
\end{eqnarray}
where the $L$'s are given in terms of the function $\Ga$ by
\begin{eqnarray}
\label{L0}
L_0(\la^2) \, &=& \, {3\la^2\over 2}\,, \\
L_2(\la^2) \, &=& \, {9\la^2\over 2}\left[1-\Ga(\la^2)
-4{c_0\over M^2}\la^2\right]-{3\la^2\over 2}\,, \\
\label{L4}
L_4(\la^2) &=& {9\la^2\over 2}\left[2{c_0\over M^2}\la^2
+{5\over 6}\Ga'(\la^2)\la^2\right]\,, \\
L_6(\la^2 )\, &=& \, {9\la^2\over 2}\left[\Ga(\la^2)-1
-\Ga'(\la^2) \la^2\right]\, ,
\label{L6}
\end{eqnarray}
and the $C$'s can be read off from $L$'s:
\begin{eqnarray}
\label{C0}
C_0 \, &=& \, L_0+L_2+L_4+L_6\,, \\
C_2 \, &=& \, -\2\LF L_2+2L_4+3L_6\RF \,, \\
C_4 \, &=& \, {1\over 16}\LF L_2+6L_4+15L_6\RF\, ,
\label{C6}
\end{eqnarray}
and now we can match them with the coefficients  coming from  the 
stress energy tensor. 

Near the bounce point ($t=0$ and $a=1$ by convention), we can expand the function $\rho(t)$ in a Taylor series:
\be
\rho(t) \, = \, \left.\rho\right|_{t=0} + 
\left.{d\rho\over dt}\right|_{t=0}t + 
\left.{1 \over 2}{d^2\rho\over dt^2}\right|_{t=0}t^2 + 
\left.{1 \over 6}{d^3\rho\over dt^3}\right|_{t=0}t^3 +
\left.{1 \over 24}{d^4\rho\over dt^4}\right|_{t=0}t^4 + \dots
\ee
The terms with odd powers of $t$ vanish by our assumption of a symmetric
bounce with bounce point $t=0$. Replacing time derivatives by derivatives
with respect to the scale factor, we obtain
\be
\rho(t) \, \approx \, 
\left.\rho\right|_{a=1} + 
\left.{\la^2t^2\over 4}{d\rho\over da}\right|_{a=1} +\left.{\la^4t^4\over 96}\LF{d\rho\over da}\right|_{a=1}+\left.3{d^2\rho\over da^2}\right|_{a=1}\RF \, ,
\label{rho}
\ee
where we have used the specific ansatz (\ref{cosh}).

Thus, by matching the $C$ 's in (\ref{Gseries}) with the coefficients of the stress energy tensor (\ref{rho}), we essentially have three equations which 
determine the two dynamical quantities, $\rho_b,\la$ along with giving us a  
constraint. One can check that the exact solutions that were found in \cite{BMS} in the presence of a 
cosmological constant and radiation is a subset of the approximate 
solutions described above. 

In the case when one can apply the ideal fluid approximation (\ref{state}), 
the equations simplify considerably and one obtains 
\be
-{3\la^2(12{c_0\over M^2}\la^2+\Ga'\la^2)\over 8}
\, = \, {2\la^2\left[6\Ga-7-4\Ga' \la^2\right]\over 1+\om}
\, = \, {48\la^2\left[42\Ga-43-30\Ga' \la^2
+24c_0\la^2\right]\over (1+\om)(9\om +11)}
\, = \, {\rho_b\over M_p^2}\,.
\ee
determining $\la^2,\rho_b$ and giving rise to an additional constraint. One can rewrite the above equations in a more illuminating form:
$$
 {\cal K}\LF{\la^2\over M_{\ast}^2}\RF=0
$$
$$
1+\om \, = \, {16{\cal F}({\la^2\over M_{\ast}^2})\over 3\la^2{\cal G}\LF{\la^2\over M_{\ast}^2}\RF}
$$
and 
$$
\rho_b \, = \, {3M_p^2\la^2{\cal G}({\la^2\over M_{\ast}^2})\over 8} \, 
$$
where we have defined dimensionless functions:
\be
{\cal F}\LF{\la^2\over M_{\ast}^2}\RF\equiv 6\Ga-7-4\Ga' \la^2
\ee
\be
{\cal G}\LF{\la^2\over M_{\ast}^2}\RF\equiv -\LF 12{c_0\la^2\over M^2}+\la^2\Ga'\RF
\ee
and
\be
{\cal K}\LF{\la^2\over M_{\ast}^2}\RF\equiv {942\Gamma-955-676\Gamma'\lambda^2
+576 {c_0\lambda^2\over M^2}\over 36 {c_0\lambda^2\over M^2}
-61\Gamma'\lambda^2+96\Gamma-112 }-3{{\cal F}\LF{\la^2\over M_{\ast}^2}\RF\over{\cal G}\LF{\la^2\over M_{\ast}^2}\RF } \,
\label{K}
\ee

A few comments are now in order. Since we want to restrict ourselves only to matter sources which are 
``consistent'' (for instance ghost-free), they have to satisfy the 
dominant energy condition. This, in particular, implies
\be
\rho(1) \, > \, 0\mx{ and } \left.{d\rho\over da}\right|_{a=1} \,< \, 0 \, \im -1\leq \om \leq 1 
\label{constraints}
\ee
imposing crucial restrictions on when one can have such a bounce 
and therefore prevent the singularity. Specifically, it says that  the dimensionless functions that we introduced, viz. ${\cal F,\ G,\ K}$, has to be positive at the solution point.

One may also wonder why we need to 
keep track of terms up to  $\cO(t^4)$. Naively, one would expect that matching 
terms up to $\cO(t^2)$ should be sufficient to capture the physics near the 
bounce. We should remember that, for consistency, we need to solve the full 
Einstein equations, \ie $\ti{G}_{mn}=T_{mn}$ up to $\cO(t^2)$. In terms 
of the ``Hubble equation'' (\ref{tildeG00}) this is equivalent to looking 
at terms  up to $\cO(t^4)$. 
\section{Modeling the transition from Hagedorn to Radiation}\label{transition}

We have seen in our analysis that for phenomenological success of the model, the bounce point should occur when string scale energy density is reached, \ie right around the 
time when the massive modes are being excited and we are entering (or just 
entered the Hagedorn phase). In fact this is also important  for the consistency of our computations of the perturbation spectrum, 
as we now illustrate. In order to be able to do any kind of ``local'' analysis 
with the metric fluctuations it is imperative that we are able to define a local matter density 
function, $\rho(x)$, which is sourcing the metric. Now this is possible 
if the following limit is well defined:
\be
\rho(x) \, = \, \lim_{r\ra 0}{M_r\over V_r} \, ,
\label{limit}
\ee 
where $M_r$ and $V_r$ are the mass/energy and volume respectively inside 
a sphere of radius $r$ around the point $x$. It is clear that a non-trivial 
and a well-defined  limit only exists iff $M_r\sim r^3$ which is indeed the 
case for ordinary particles. However in the Hagedorn phase $E\sim r^2$ and therefore the density function 
is ill defined and accordingly one cannot proceed to obtain its Fourier 
transform or power spectrum for that matter. Thus in order for us to perform 
the usual density perturbation
analysis it is imperative that for the distance scales that are relevant for 
CMB, the energy is still dominated by the contribution from the  massless 
modes which would give rise to an $r^3$ term\footnote{As long as our 
detectors are sufficiently coarse 
grained, we will then have a well defined limit (\ref{limit})}.

Now, in order to check whether this is true one would need to compare 
the energy in  the massive string modes (Hagedorn matter) given by (\ref{temperature})\footnote{Actually, from (\ref{temperature}) one finds $E_{\mt{hag,total}}=E_{\mt{massive}}+k(M_s r)^3$. However, the second term does not contribute to the fluctuations (heat capacity), neither does it pose any problem for the existence of the limit (\ref{limit}), since it scales as volume. In fact one sees that already the full expression of the Hagedorn energy suggests a regime ($E_{\mt{massive}}\ll k(M_s r)^3$) when the fluctuations come from $E_{\mt{massive}}$, but the energy is still dominated by the term linear in volume. Perhaps, this term could even be interpreted as the contribution from the massless radiative modes at the transition point. Unfortunately, we do not have a clear understanding of the thermodynamics during this transition phase and hence we had to ``introduce'' radiation by hand to capture the physics. Hopefully future research will clarify the situation further.}
\be
E_{\mt{massive}} \, = \, c_{\mt{hag}}r^2M_s^3\left\{\ln\LT{T\over 1-T/T_H}\RT+\log\log \mx{ terms}\right\}
\ee
with that of  radiation (massless modes)
\be
E_{\mt{rad}} \, = \, c_{\mt{rad}}T^4r^3
\ee
inside spheres of the relevant scales. Here 
$c_{\mt{rad}},c_{\mt{hag}}\sim {\cal O}(1)$ constants.  Now, in order for 
the limit (\ref{limit}) to exist, minimally we require
\be
\zeta_E \, \equiv \, {E_{\mt{rad}}\over E_{\mt{massive}}}
\, \sim \, 
\LF{T\over M_s}\RF^4 {M_s r_0\over \ln\LT{T\over 1-T/T_H}\RT}
\, > \, 1 \, ,
\label{energy-ratio}
\ee
where $r_0$ is the physical scale at the time of the bounce corresponding to 
the Hubble radius today. Therefore as long as the logarithm in the denominator $\sim \cO(1)$, since at the time of the bounce $T\sim M_s$, to satisfy (\ref{energy-ratio}) we simply require
\be
\zeta_E \, \sim \, M_s r_0 \, > \, 1 \, .
\ee

It is easy to estimate this quantity. We know that during the bounce the 
universe expands very little. Therefore,  making  the simplifying assumption 
that most of the expansion of our universe occurred during the radiation era 
since string scale energy density, we have 
\be
{r_0\over H_0^{-1}} \, = \, {T_0\over M_s}\Ra M_s r_0
\, \sim \, {M_p\over T_0}\sim 10^{30} \, \gg \, 1 \, .
\ee
Indeed, (\ref{energy-ratio}) is easily satisfied for the relevant scales 
that we are observing in CMB today and therefore our analysis of perturbation 
spectrum is well justified. As an aside, we notice that for all scales $r>r_0$ 
the energy is going to be 
dominated by radiation and in particular the evolution of our universe would 
also be governed by  radiation.

What now becomes crucial is to verify that it is still the massive string 
modes of the Hagedorn phase that dominate the energy fluctuations in a 
sphere of radius $r_0$, and not the massless modes. The amplitudes of 
fluctuation is proportional to the respective specific heats (more precisely the heat capacities) and therefore 
in order for the power spectrum (\ref{power}) to hold we need
\be \label{ineq2}
\zeta_C \, \equiv \, {C_{\mt{rad}}\over C_{\mt{massive}}} \, < \, 1 
\ee    
Differentiating the expressions for the energies with respect to 
temperature for the respective phases we have
\be
\zeta_C \, = \, r_0 M_s \bigl( 1 - {T \over {T_H}} \bigr) \, .
\ee
Thus, in order to satisfy (\ref{ineq2}) one requires
\be
{\De T\over T_H} \, \equiv \, 1-{T\over T_H} \, < \, 10^{-30} \, .
\ee
\section{Including the Dilaton and Other Light Moduli Fields}\label{moduli}

So far we have not discussed the possible role of moduli fields, such
as the dilaton or radion that are present in any string theory
compactifications.  It is well-known that for consistency of late time
physics (fifth force constraints, and bounds on variation of physical
constants) one typically requires\footnote{It is possible construct some string theory motivated scenarios where one could avoid these observational constraints and still have a rolling light moduli field playing the role of a quintessence matter \cite{tuomas},  which has become attractive due to the recent observations of dark energy.} that these moduli fields are stabilized at a scale
which is at least higher than the scale of big bang nucleosynthesis,
\ie Mev. There are two different possibilities that are compatible
with our scenario. (i) If the scale of stabilization is higher than
the string scale, then the bounce never probes this regime and the
moduli fields always remain frozen playing no role either in the
background evolution or in the generation of fluctuations. (ii) A
second possibility, that the moduli are stabilized at a scale much
lower than the string scale where all the dynamics is taking place is
also consistent with our scenario. Let us look at this more carefully.

In the Einstein frame, the action for these scalars  just looks like a 
free theory
\be
S_{\phi} \, = \, -\2\int d^4x \sqrt{-g}(\p\phi)^2 \, ,
\label{dilaton}
\ee
where $\phi$ has been appropriately normalized.  An important subtlety is 
that after the conformal transformation (\ie in the Einstein frame) the 
moduli fields such as the dilaton couple to the massive string modes. 
Thus, if there is stringy Hagedorn matter (massive closed string modes) present, then such matter
would typically 
source the dilaton. However, as we explained before, we are primarily 
interested in a scenario where the energy density is not yet dominated 
by Hagedorn matter (only the fluctuations are) and in fact its energy density
is hugely suppressed as compared to radiation. Therefore, such a source term 
can be completely ignored. In this context we note that since radiation is 
conformal, it does not yield any coupling to the dilaton after the conformal
transformation. Let us therefore investigate the evolution of the free theory (\ref{dilaton}). 

Once we include the dilaton, the generalized Einstein equation reads
\be
\ti{G}_{00} \, = \, {1\over M_p^2}(\rho_{\mt{rad}}+\rho_{\phi})
\ee  
where
\be
\rho_{\phi} \, = \, {\dot{\phi}^2\over 2} \, .
\ee
This has to be supplemented by the dilaton equation
\be
\ddot{\phi}+3H\dot{\phi}\, = \, 0 \, .
\ee
 
One can solve these equations. For instance, the dilaton equation 
yields
\be
\rho_{\phi} \, = \, {\dot{\phi}^2\over 2}=\rho_{\phi,b}a^{-6} \, .
\ee
Using the specific bounce solution we find
\be
\phi(t) \, = \, \phi_b+\sqrt{\rho_{\phi,b}\over M_p^2\la^2}\LT \tanh(t\sqrt{\la/2})\mt{sech}(t\sqrt{\la/2}) +2\tan^{-1}(e^{t\sqrt{\la/2}})-{\pi\over 2}\RT \, .
\ee
It is interesting to note that during the course of the bounce $\phi$ 
changes only by a finite amount:
\be
\De \phi\equiv \phi(\infty)-\phi(-\infty) \, = \, \pi\sqrt{\rho_{\phi,b}\over M_p^2\la^2}\, .
\ee

Thus, we observe that if
\be
{\rho_{\phi,b}\over M_p^2\la^2}\, \ll \, 1 \, ,
\ee
the string coupling 
\be
g_s \, \sim \, e^{\phi/M_p}
\ee 
changes very little. Moreover, since the kinetic energy of the dilaton is 
small, it  does not effect the evolution of the metric. In particular, it is 
completely consistent to set the value of $\phi$ to be a constant.


\end{document}